\newcommand{\SCB}{SrCu$_2$(BO$_3$)$_2$\xspace}

\newcommand{\be}{\begin{equation} }
\newcommand{\ee}{\end{equation} }
\newcommand{\bea}{\begin{eqnarray} }
\newcommand{\eea}{\end{eqnarray} }

\documentclass[aps,prl,reprint,showpacs,longbibliography]{revtex4-1}
\usepackage{fullpage}
\usepackage{wrapfig}
\usepackage{floatflt}
\usepackage{xspace}
\usepackage{dsfont}
\usepackage{amsmath,amssymb,bm}
\usepackage{graphicx}
\usepackage{xspace}
\usepackage{latexsym}
\usepackage{color}
\usepackage{hyperref}
\usepackage{placeins}
\usepackage[utf8]{inputenc}
\begin{document}
\title{Sign-switching of dimer correlations in \SCB under hydrostatic pressure}

\author{S.~Bettler}
    \email{sbettler@phys.ethz.ch}
\affiliation{Laboratory for Solid State Physics, ETH Z\"{u}rich, 8093
Z\"{u}rich, Switzerland}

\author{L.~Stoppel}
    \affiliation{Laboratory for Solid State Physics, ETH Z\"{u}rich, 8093
Z\"{u}rich, Switzerland}

\author{Z.~Yan}
    \affiliation{Laboratory for Solid State Physics, ETH Z\"{u}rich, 8093
Z\"{u}rich, Switzerland}

\author{S.~Gvasaliya}
\affiliation{Laboratory for Solid State Physics, ETH Z\"{u}rich, 8093
Z\"{u}rich, Switzerland}

\author{A.~Zheludev}
    \email{zhelud@ethz.ch}
    \homepage{http://www.neutron.ethz.ch/}
    \affiliation{Laboratory for Solid State Physics, ETH Z\"{u}rich, 8093
Z\"{u}rich, Switzerland}

\begin{abstract}
	Magnetic and vibrational excitations in \SCB are studied using Raman spectroscopy at hydrostatic pressures up to 34~kbar and temperatures down to 2.6~K. The frequency of a particular optical phonon, the so-called pantograph mode, shows a very strong anomalous temperature dependence below about 40~K. We link the magnitude of the effect to the magnetic exchange energy on the dimer bonds in the Sutherland-Shastry spin lattice in this material. The corresponding dimer spin correlations are quantitatively estimated and found to be strongly pressure dependent. At around $P_2\sim 22$~kbar they switch from antiferromagnetic to being predominantly ferromagnetic. 
\end{abstract}

\date{\today}
\maketitle
\begin{figure}[h!]
\centering
\includegraphics[width=\columnwidth]{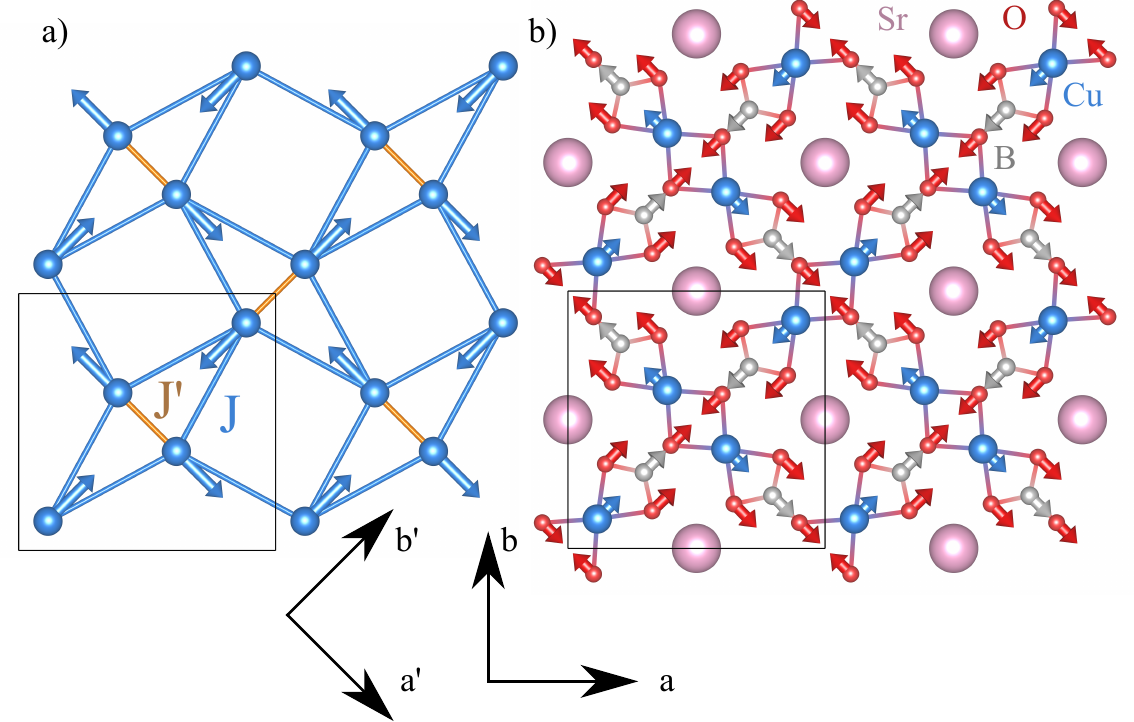}
\caption{a) Lattice of orthogonal coupled dimers in \SCB with intra-dimer interaction J' and inter-dimer interaction J. Copper atomic motions of the pantograph mode are indicated as arrows. b) Relative atomic displacements of the pantograph mode in a single layer of \SCB.}
\label{fig::pantograph}
\end{figure}

The Shastry-Sutherland model (SSM) is arguably one of the most important
constructs in the field of quantum magnetism.
It demonstrates that a gapped quantum paramagnet can occur
in a well-connected Heisenberg spin Hamiltonian beyond the
unique topology of a single dimension.  Its key feature is geometric
frustration of antiferromagnetic (AF) interactions
between AF $S=1/2$ dimers arranged on a particular 2-dimensional lattice
(Fig.~\ref{fig::pantograph}a) \cite{Shastry1981}.
For sufficiently strong  frustration the {\em exact} ground state is a
product of AF singlets on each dimer bond $J'$. For weak frustration
one recovers the semi-classical Néel-ordered phase.  What happens
in between has been hotly debated \cite{Albrecht_1996,Miyahara99,Weihong99, Koga99,MuellerHartmann2000,Takushima2001,Carpentier2001,Chung2001,Weihong2001, Lauschli2002,Hajj2005, Darradi2005, Isacsson2006,Moukouri2008,Moliner2011, Corboz2013,Ronquillo2014,wang2018,boos2019}.
The most intriguing intermediate phase proposed is
the so-called plaquette state \cite{Lauschli2002,Takushima2001,Koga99}. Dimer singlets are destroyed to be replaced by singlets
composed of {\em  four} spins connected via the inter-dimer bonds $J$.
Translational symmetry is broken and some or all of the dimer spin correlations become predominantly {\em ferromagnetic} (FM).

The only known and much studied experimental realization of the model is
\SCB, where $S=1/2$ Cu$^{2+}$ ions form dimers via Cu-O-Cu superexchange
pathways, which are connected through (BO$_3$) units (Fig.~\ref{fig::pantograph}b) \cite{Kageyama1999}. With a frustration ratio $J/J'\sim 0.6$ \cite{Knetter2004} it is reliably in the dimer phase, with a spin gap $\Delta=3$~meV \cite{Kageyama1999} in
the excitation spectrum. We are incredibly lucky that in this material
$J/J'$ can be continuously tuned by hydrostatic pressure \cite{zayed2017}. The frustration
ratio increases steadily with pressure, eventually leading to a Néel
ordered state above 30~kbar \cite{guo2019}.  Moreover, already at $P_c\sim 18$~kbar the original dimer phase gives way to a new quantum paramagnet, presumed to be
the plaquette state \cite{Haravifard2012,Waki2007,zayed2017, Sakurai2018,guo2019}. Thus, theoretical predictions for exotic phases of the SSM are put to the experimental test.

How can one be sure that the novel phase is indeed plaquette-, rather than
dimer-based? To date, the only supporting evidence comes from studies of
the wave vector dependence of inelastic neutron scattering intensities.\cite{zayed2017}
Performing such measurements in a bulky cell needed to produce the required
pressure for a sufficiently large sample is a formidable task. The
resulting data are unavoidably limited and noisy, leaving the
interpretation depending on strong assumptions and theoretical modeling \cite{zayed2017}. In
the present Rapid Communication we use an entirely different approach. We infer the
strength of dimer spin correlations in \SCB from their effect on certain
optical phonons, which can be measured using Raman spectroscopy in a
diamond-anvil pressure cell. We show that around $P_c$ these correlations
switch from AF to dominantly FM, and thereby independently confirm the destruction of the AF-dimer ground state. Moreover, we obtain a {\em quantitative}
estimate for the dimer spin correlations at pressures up to 34~kbar.

The main idea is as follows. As has been established in other dimer
systems, the development of pair spin correlations at low temperatures
leads to a magnetic contribution to the rigidity of the corresponding bond \cite{Raas2002,Choi2005,Bettler2017}.
This, in turn, gives rise to anomalous shifts of certain phonon
frequencies at unusually low temperature.
In complex structures it may be difficult to associate these shifts with any
particular magnetic bonds \cite{Bettler2017}. For the highly symmetric structure of \SCB
though, the assignment of measured phonons to particular atomic motions is greatly simplified. At $\hbar \omega\sim$198~$\mathrm{cm}^{-1}$ (=24.5~meV)
there is a specific optical excitation, the so-called pantograph mode,
visualized in Fig.~\ref{fig::pantograph}b) \cite{Radtke2015}.

It directly modulates the  the intra-dimer coupling constant $J'$, which is especially sensitive due to an almost 90$^\circ$ Cu-O-Cu \cite{Radtke2015}.
At same time, its effect on $J$ is expected to be negligible in comparison\cite{Vecchini2009,Radtke2015}.
The frequency of this phonon is related to the second derivative of the energy
of the corresponding lattice distortion with respect to the distortion amplitude $u$:  $\omega^2\propto \partial^2E/\partial u^2$.
Each dimer's magnetic contribution is  $J'\mathbf{S}_1 \mathbf{S}_2$.
The frequency of the pantograph mode is much larger than the magnetic energy scale.
This ensures  an
adiabatic coupling scenario where the spin correlations are unaffected by the phonon which is ``too fast''.
Since the vibration mode spans the entire sample, the magnetic energy needs to be summed (averaged) over all dimers. 
With this in mind:
\be
\delta \omega=\omega-\omega_0\propto \omega^2-\omega_0^2 \propto\frac{\partial^2 J'}{\partial u^2} \, \langle \mathbf{S}_1 \mathbf{S}_2 \rangle,\label{emag}
\ee
where $\omega_0$ is the mode frequency in the absence of any magnetic correlations and it is assumed that $|\omega-\omega_0|\ll \omega_0$.
The anomalous magnetic frequency shift of the pantograph mode is thus proportional to the dimer spin correlator. This
also follows from the calculation of the corresponding spin-phonon coupling constant \cite{Choi2003} that assumes  a particular power-law $u$-dependence of $J$ \cite{Miyahara2003}.

\begin{figure}[h!]
	\centering
	\includegraphics[width=\columnwidth]{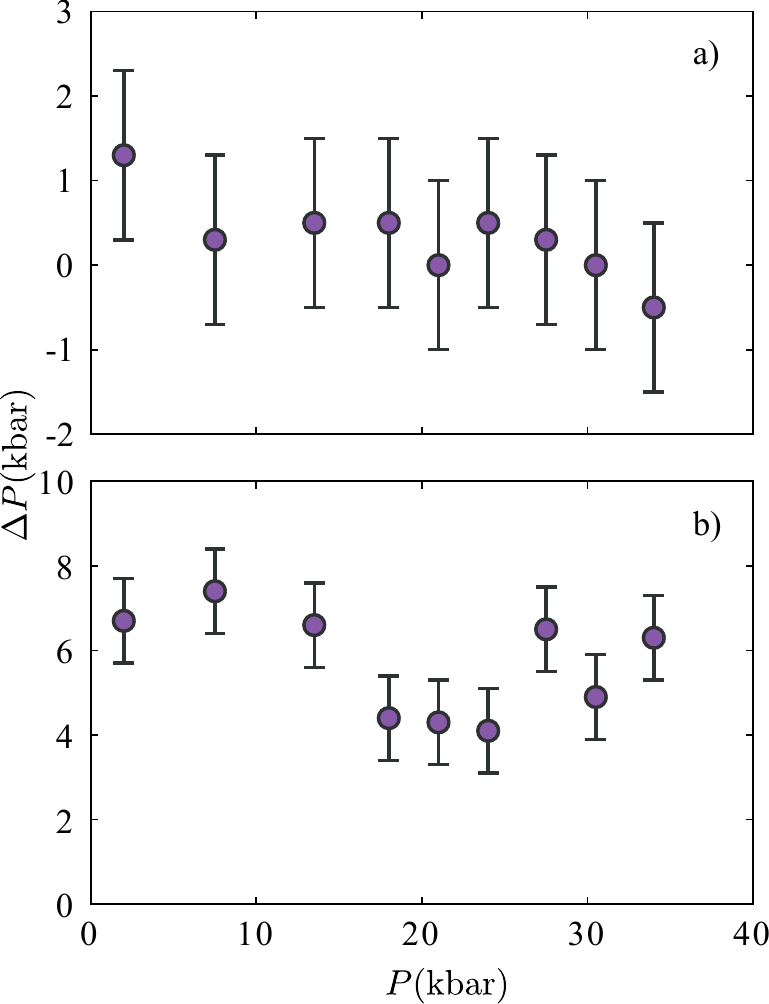}
	\caption{a) Measured pressure changes between 40K and base temperature. b) Measured pressure changes between room temperature and 40K. }
	\label{sup2}
\end{figure}

Optical phonons and magnetic
excitations have been extensively studied in \SCB using Raman spectroscopy
at ambient pressure\cite{Lemmens2000,Choi2003,Gozar2005,Homes2009}. In the present work we perform similar
measurements in a 1-mm-culet diamond-anvil cell  at pressures up to 34~kbar
using a Trivista 557 triple-grating spectrometer and a liquid-nitrogen-cooled CCD detector. The data were collected in a backscattering geometry
using a focusing microscope. The incident laser wavelength was 532 nm. The cell environment was a He-flow cryostat with
base temperature of 2.5~K.
The pressure transmitting medium was argon. The pressure was determined in-situ
from the shift of the ruby R1 fluorescence line\cite{Feng2010, Zha2000}. The temperature dependence of the effect, which becomes particularly important above 50~K, was explicitly corrected for \cite{Ragan1992}. The pressure changes below 40~K are marginal (Fig. \ref{sup2}a). Most of the pressure change due to thermal expansion and solidification of argon occurs between room temperature and 40~K (Fig. \ref{sup2}b).

\begin{figure}[h!]
\includegraphics[width=\columnwidth]{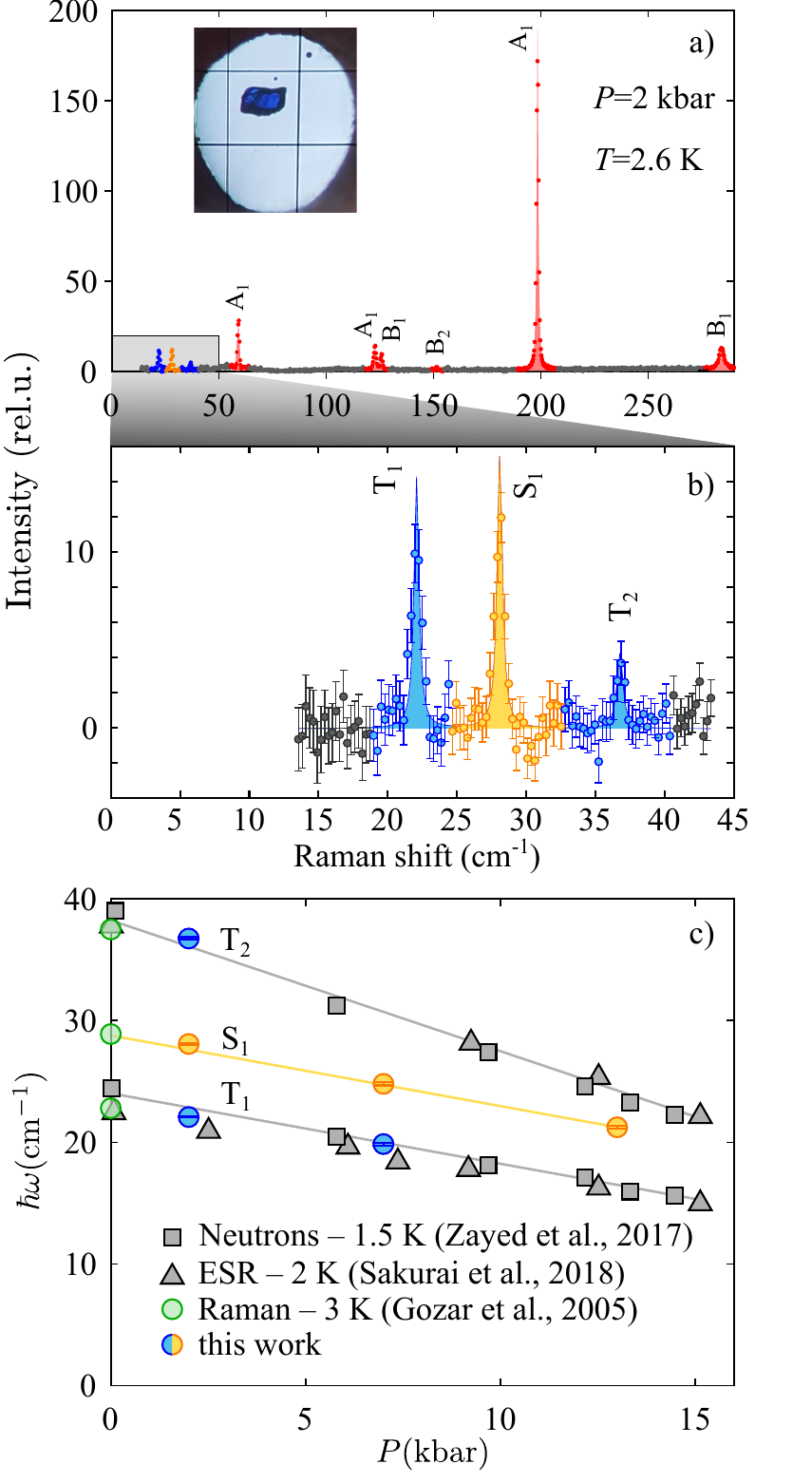}
\caption{a) Raman data at T$\approx2.6~$K and 2~kbar in $\bar{c}(a'b')c$ polarization measured with $\lambda$=532~nm laser. Symmetries of phonons are assigned as in Ref.~\onlinecite{Gozar2005}. Inset: picture of the single crystal sample inside the pressure cell along with two ruby spheres used for pressure calibration. b) Blowup of the low-energy portion of a). Triplets $T_1$ and $T_2$ and singlet $S_1$ are assigned according to Ref~.\onlinecite{Gozar2005} c) Comparison of excitation energies of singlet and triplet excitations to published results \cite{Sakurai2018,Gozar2005,zayed2017}.}
\label{fig::spectra}
\end{figure}

 All pressure values quoted below correspond to base temperature readings.
All experiments were performed on a $\sim$0.2$\times$0.1$\times$0.05 mm$^3$ single crystal sample extracted from cleaving a crystal grown using the floating-zone technique. The incident light was perpendicular to the \emph{ab} cleavage plane (see inset in Fig.~\ref{fig::spectra}). Unless otherwise noted, all spectra shown were measured in the $\bar{c}(a'b')c$ scattering geometry. An independently measured background from the pressure transmitting medium was subtracted from all spectra. 

Optical
measurements on \SCB are extremely challenging due to the narrow band gap
of the material, which leads to high absorption\cite{Homes2009}. The biggest problem is
unwanted heating of the sample by the incident laser beam. To suppress this
effect, the data were collected at very low power, as low as  0.05~mW at the
lowest temperatures. As a result, typical counting times were as much as
72~h.  In all cases it was verified that further reducing the power had no
effect on the resulting spectra.
A typical low-energy Raman spectrum collected in \SCB at a pressure of
$P=2$~kbar and $T=2.6$~K is shown in Fig.~\ref{fig::spectra}a). The observed frequencies appear fully consistent with previous measurements at ambient pressure \cite{Gozar2005}.
The three visible lowest-energy excitations are magnetic in origin: two
triplets and one singlet. All remaining peaks are phonons. The peak widths are in all cases below 1~cm$^{-1}$ and correspond to the expected instrumental resolution.
Since data for the entire spectrum are accumulated concurrently, this confirms that temporal drift of spectrometer alignment during the long data collection period is a non issue.

The pantograph
mode is the peak at around 198~cm$^{-1}$. It was identified by a density functional theory (DFT) ab-initio  calculation with the Quantum Espresso software package \cite{QUANTUM_ESPRESSO_2009}, using the SSSP Accuracy (version 1.1) pseudopotential library \cite{prandini2018precision,Lejaeghere2016,Hamann2013,SCHLIPF2015,prandini2018,GARRITY2014,DALCORSO2014}. The kinetic energy cutoff was 120~Ry and the charge density cutoff 600~Ry. Brillouin zone integration was performed using a 4$\times$4$\times$4 k-point grid. For better convergence a Marzari-Vanderbilt cold smearing of 0.01~Ry was applied.

In order to validate the experimental methodology we first checked the
behavior of the three {\em magnetic} excitations. With increasing pressure
all three peaks shift to lower energies and progressively weaken. Their
measured frequencies are plotted in colored circles in
Fig.~\ref{fig::spectra}c). At higher pressures they become undetectably weak
or shift outside our measurement window ($T_1$). The observed softening of
the triplet modes is fully consistent with previous Electron spin resonance(ESR) \cite{Sakurai2018} and inelastic
neutron scattering \cite{zayed2017} studies, shown as triangles and squares in the figure.
The softening of the singlet mode is a new result, since neither
ESR nor neutrons are sensitive to singlet-singlet transitions.
\begin{figure}
\centering
\includegraphics[width=\columnwidth]{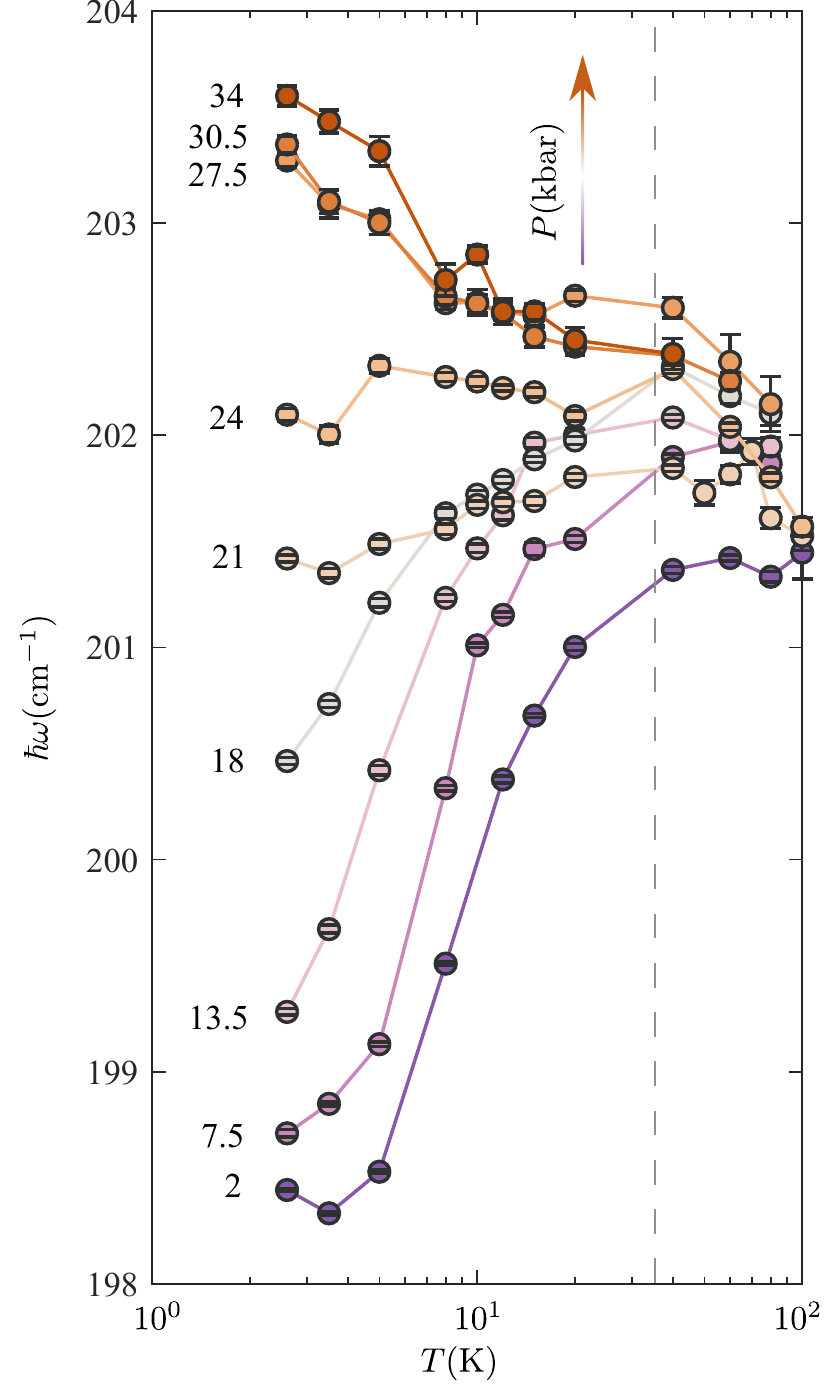}
\caption{Measured temperature dependence of the pantograph mode frequency in \SCB at different pressures. Note the logarithmic temperature scale. The dashed line is the temperature corresponding to the spin gap energy at ambient pressure.}
\label{fig::vsT}
\end{figure}

%=========================================================================================

The central result of this Rapid Communication is the observation of an anomalous
temperature dependence of the pantograph mode, as visualized in
Fig.~\ref{fig::vsT}. Note the logarithmic temperature scale. At
all pressures, the pantograph mode undergoes at most a modest hardening upon cooling
down to 40~K. This behavior is  due to the usual
anharmonicities of lattice vibrations and the temperature dependence of pressure in the loaded cell at higher temperatures. 

\begin{figure}[h!]
	\centering
	\includegraphics[width=\columnwidth]{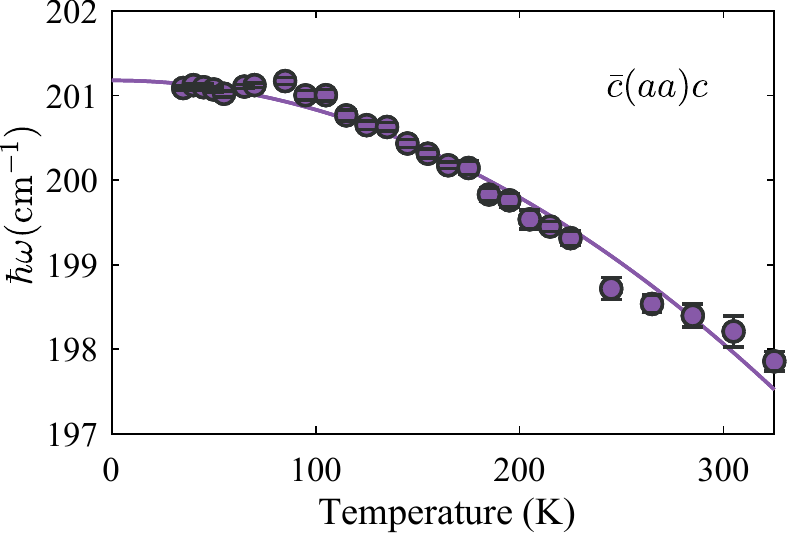}
	\caption{Measured temperature dependence of the pantograph mode in \SCB at ambient pressure (symbols) and fit assuming anharmonic multi-phonon coupling (solid line).}
	\label{sup1}
\end{figure}

At temperatures below 40~K the anharmonic thermal frequency shift of the pantograph mode is negligible. In order to establish this, we measured the temperature dependence of the pantograph mode frequency between 35~K and room temperature at ambient pressure. The result is visualized in Fig.~\ref{sup1} (symbols). These data were analyzed using the well-known predictions for anharmonic multi-phonon coupling (Eq.~3.9 in \cite{balkanski1983}). An excellent fit is obtained using empirical parameters $\omega_0= 201.2~\mathrm{cm}^{-1},\,C=0$ and $D=-2.7(1)\cdot 10^{-3}~\mathrm{cm}^{-1}$ (solid line in Fig.~\ref{sup1}). Extrapolating this curve to below 40~K yields an estimate of the anharmonic shift in the low-temperature range: $\delta\omega\lesssim 0.1$~cm$^{-1}$. This is considerably smaller than the observed anomalous magnetic frequency shift and can be safely disregarded.

\begin{figure}
\centering
\includegraphics[width=\columnwidth]{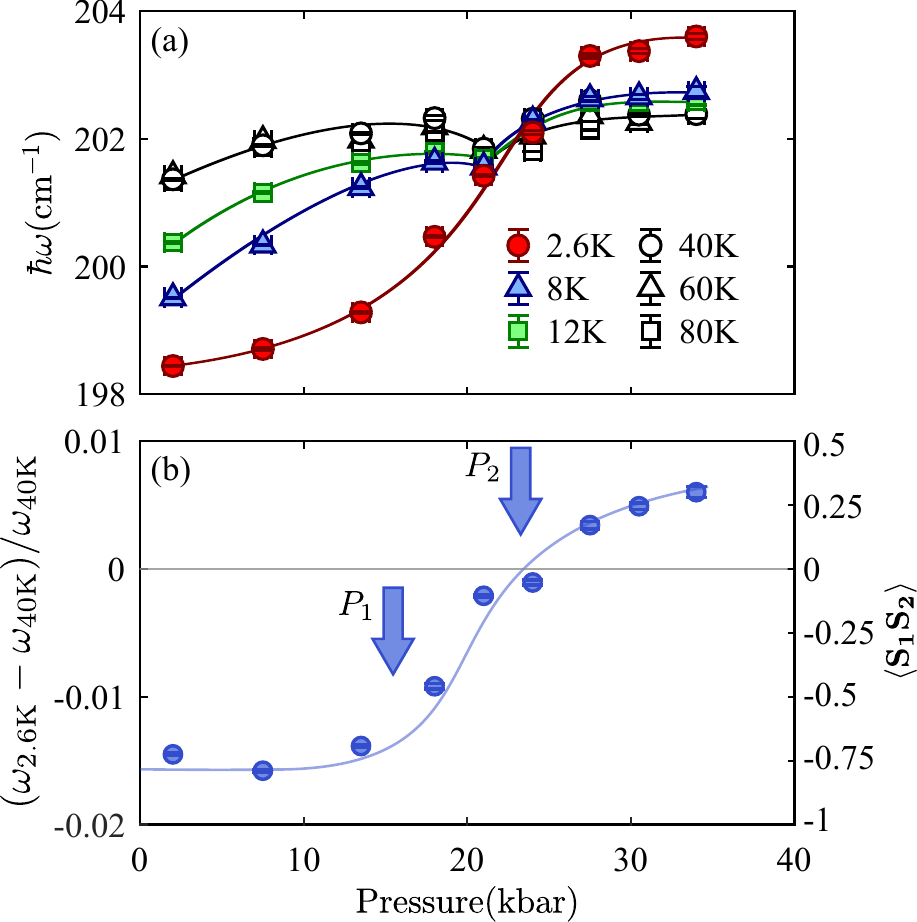}
\caption{a) Measured pressure dependence of the pantograph mode frequency in \SCB at several different temperatures (symbols). The lines are guides for the eye. b) Pressure dependence of the low-temperature frequency shift of the pantograph mode. $P_1$ indicates the beginning of the destruction of dimer correlations and $P_2$ denotes the sign switching of dimer correlations.  The solid line is a guide to the eye.}
\label{fig::vsPP}
\end{figure}

Below this point the mode suddenly becomes strongly $T$- dependent.
At low applied pressures it{\em softens} upon cooling to the base temperature by as much as 1.5\%. At the highest
pressures the effect is reversed: in the same low-temperature interval the excitation hardens by as much as 0.5\%. 
Also revealing is the pressure dependence at different temperatures shown in Fig.~\ref{fig::vsPP} a).
For $T\gtrsim 40$~K the pressure dependencies are almost $T$-independent. They all show a monotonous hardening with a small but reproducible dip at about $20$~kbar. 
At lower temperatures the situation changes drastically, the frequency showing a steep almost step-like increase. 
%
%\begin{figure}
%	\centering
%	\includegraphics[width=\columnwidth]{pantograph_softening_pdep_smooth.pdf}
%	\caption{ Measured pressure dependence of the pantograph mode frequency in \SCB at several different temperatures (symbols). The lines are guides for the eye.}
%	\label{fig::vsP}
%\end{figure}

As established above, at temperatures below 40~K anharmonic effects extrapolated from measurements above $35$~K are entirely negligible. The only remaining relevant energy scale is the magnetic one.
We conclude that the anomalous temperature dependence of the pantograph mode
is due to magneto-elastic coupling. 
The corresponding relative frequency shift between 40~K and 2.6~K is plotted against pressure in Fig.~\ref{fig::vsPP} b)
(left axis). As mentioned above, the cell pressure is practically constant in this temperature range.
The shift remains flat up to about $P_{1}\sim 15$~kbar and then
steadily increases up to the highest attainable pressures, switching sign
at about $P_{2}\sim 22$~kbar. 

 As shown by ESR experiments \cite{Sakurai2018},  $J'$ decreases by only about 12\% from 0 to 15~kBar. This allows us to assume that 
the 2nd derivative in Eq.~\ref{emag} is also {\em only weakly pressure-dependent} to some approximation.
The lack of any pressure dependence for $P\lesssim P_1$  in Fig.~\ref{fig::vsPP} b) confirms this. Here the system
is clearly in the dimer phase \cite{zayed2017}. In this regime the exact result is $ \langle \mathbf{S}_1 \mathbf{S}_2 \rangle
\equiv -\frac{3}{4}$ regardless of $J'/J$ \cite{Miyahara99}, and the measured constant frequency shift implies a constant ${\partial^2 J'}/{\partial u^2}$. 
Assuming the trend continues at higher pressures, to within a scale factor Fig.~\ref{fig::vsPP} b)  represents
the pressure dependence of $\langle
\mathbf{S}_1 \mathbf{S}_2 \rangle$. The exact value for the dimer phase  provides a
calibration for the entire plot  (Fig.~\ref{fig::vsPP}b), right axis). At $P_2$ the relative
alignment of nearest neighbor spins {\em switches sign and becomes predominantly
	ferromagnetic}.

 That AF dimers are replaced by a completely new spin correlation pattern already at rather low pressures is significant. The only other indication of that is the rather tenuous 22~kbar strucure factor data from the neutron study of Ref.~\cite{zayed2017}. The transformation that we observe starts already at about 15~kbar. It can only be driven by a subtle change in the $J'/J$ ratio, as assumed in most theoretical studies \cite{Albrecht_1996,Miyahara99,Weihong99, Koga99,MuellerHartmann2000,Takushima2001,Carpentier2001,Chung2001,Weihong2001, Lauschli2002,Hajj2005, Darradi2005, Isacsson2006,Moukouri2008,Moliner2011, Corboz2013,Ronquillo2014,wang2018,boos2019}.  It takes place long before the exchange constants themselves can be expected to switch sign \cite{Sakurai2018}. It also seems not to be associated with any structural transition, of which we do not observe any obvious signs such as splitting of phonon lines or the appearance of new modes.

	According to a recent calorimetric study \cite{guo2019}, there is a distinct thermodynamic phase that emerges below $T_c\sim 2$~K  just about $P_1$. The authors suggest that it may be a plaquette state. The sign-switching reported here is consistent with that interpretation. While in the dimer phase the spin correlations are AF, in the plaquette state they at least partially switch to FM and may evolve continuously\cite{Corboz2013,boos2019}. Unfortunately, the lowest attainable temperature in our experiments is slightly above $T_c$. The observed behavior can be interpreted as a change in the character of dominant {\em short-range} spin correlations, which are just about to order in another plaquette configuration upon further cooling. According to \cite{guo2019}, N\'eel magnetic order sets in above approximately $P_2$. Here even stronger FM correlations are predicted \cite{Corboz2013}, and indeed observed in our measurements. 

In summary, our measurements of the pantograph mode in
\SCB provide an indirect but precise measurement of a pressure-induced sign switching of nearest-neighbor dimer spin
correlations. We hope that a quantitative comparison with theoretical studies will become possible in the future.

This work was partially supported by the Swiss National Science Foundation,
Division 2. We thank D. Blosser (ETHZ) for help with first principles calculations, E. Pomjakushina (PSI) for guidance on the single crystal growth and F. Mila, H. Rønnow and D. Badrtdinov (EPFL) for enlightening discussions.

\bibliography{SB_bib}

\end{document}